# Spectroscopy of van der Waals nanomaterials: Opportunities and Challenges

S. V. Mambakkam, [1] and S. Law [2*]

[1]*Department of Materials Science and Engineering, University of Delaware, Newark DE, 19711, USA*

[2]*Department of Materials Science and Engineering, Pennsylvania State University, University Park PA, 16802, USA*

*\*sal6149@psu.edu*

The study of van der Waals (vdW) materials has seen increased interest in recent years, due to the wide range of uses for these materials because of their unique mechanical, electronic, and optical properties. This area has recently expanded further into studying the behavior of vdW nanomaterials, as decreasing dimensions open up opportunities to interact with these materials in new ways. However, measuring the band structures of nanomaterials, which is key to understanding how confinement affects material properties and interactions, comes with several challenges. In this review, we survey a range of techniques for synthesizing and characterizing vdW nanomaterials, in order to outline the key material and characterization challenges. This includes controlling the Fermi level in vdW nanoparticles, preparing these particles for either ensemble or individual particle measurement, as well as protecting the pristine surface from oxidation.

## I. INTRODUCTION

Quantum-confined materials have taken on greater importance recently due to a renewed focus on quantum computing, quantum sensing, and quantum information science. Characterizing quantum-confined materials has therefore also taken on renewed importance. Understanding how the band structure of these materials depends on the material properties as well as characterizing properties of these states is foundational to quantum device design. There are a variety of phenomena which enable measurement of the band structure of quantum-confined materials including optical absorption, photoemission, electron-tunneling, and electrostatic interaction. The data obtained via these measurement techniques can be used to determine the spacing of the discrete quantum-confined states or to determine the local density of states in a quantum-confined material, and each technique has its own pros and cons.

Among the classes of quantum-confined materials attracting interest are topological materials and 2D materials [1, 2, 3, 4, 5]. Topological materials are a class of materials which host unique, topologically-protected states on their outer surface, due to a difference in the topological nature of the material relative to the surrounding medium. For instance, topological insulators, or TIs, are topological materials which host linearly-dispersed electronic states that cross the bulk band gap of the material. Due to the strong spin-orbit coupling interaction in these materials, the surface state electrons are spin-momentum locked, causing them to resist localization and scattering from non-magnetic perturbations [6, 7]. In a similar vein, 2D materials such as transition metal dichalcogenides or TMDs have also received a great deal of interest due to their unusual optical and electronic properties as well as the unusual physics that can arise when a material is confined to two dimensions [8].

In recent years, the study of these materials has pivoted to lower-dimensional structures, including nanoribbons, nanowires, nanoplates, and nanoparticles [9, 2, 10, 11, 12, 13]. Reducing the dimensions of a material produces quantum confinement in the energy states: the electron energy levels become discrete, rather than continuous, and the spacing between energy levels depends on the size of the structure. Bringing material down to 1D or 0D can lead to new physics and devices that leverage the control over the discrete energy spacing. As these nanostructures reach dimensions of 10-100 nm, theory indicates that the electronic energy level separations will typically be in the optical frequency ranges (visible through THz), which could be useful for optical sensing or quantum computation applications [14, 15, 16]. However, using these states in device applications requires both a means of synthesizing nanoparticles with a high degree of control over the dimensions and a means to probe those nanoparticles and resolve the discrete surface states. This becomes particularly challenging for nanostructures in which the electron energy separations are ~1 meV.

This article will provide an overview of methods for synthesizing and characterizing nanomaterials, with a particular focus on 2D and topological materials. We split the discussion of this subject into three sections: first, we review methods for synthesizing nanomaterials, highlighting the relative pros and cons of each synthesis technique in terms of controllability, quality of material, and so on. Next, we discuss several techniques that may be suitable for probing the band structures in nanomaterials. For each technique, we highlight key sample design requirements and considerations as it applies to measuring nanoparticles. Finally, we highlight some of the areas which require further work in this field, which involve solving specific materials challenges in both sample synthesis and characterization. This review will primarily focus on the family of chalcogenide TIs $Bi_2Se_3$, $Bi_2Te_3$ and $Sb_2Te_3$ as the model system. These compounds are all 2D topological materials, and many of the requirements for characterizing these materials are common to the characterization of any nanomaterials.

## II. SYNTHESIS METHODS

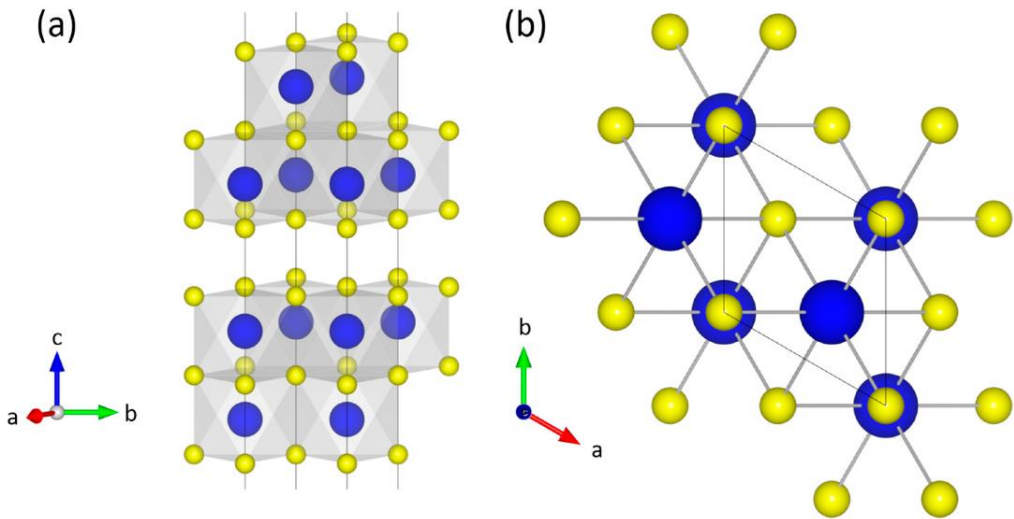

FIG. 1. Schematic of $Bi_2Se_3$ crystal structure. Larger blue atoms represent bismuth, smaller yellow atoms represent selenium. a) side profile along a-b plane showing two quintuple layers, b) top-down view along c-axis. Reprinted from Crystals, 6(11), 154 (2016), open access [17]

In this article we focus specifically on the chalcogenide TIs $Bi_2Se_3$, $Bi_2Te_3$ and $Sb_2Te_3$, which exhibit topological states delocalized on their surfaces [18]. These materials share a tetradymite crystal structure and a Dirac cone band structure. In addition, these are all 2D, or van der Waals (vdW), materials: atoms are arranged in these materials in quintuple layers (QLs) [17]. We show the example of this arrangement in Figure 1 for $Bi_2Se_3$, where the larger blue atoms represent bismuth, and the smaller yellow atoms represent selenium. Within the QL of $Bi_2Se_3$, there is a layer of selenium atoms sandwiched by two layers of bismuth atoms which are again sandwiched by two layers of selenium atoms. The atoms are strongly bonded within the QL, while each QL is bonded to the next by vdW bonds. Because of this weak bonding, applying mechanical force via techniques like exfoliation can remove individual QLs from a bulk crystal [19]. Thin films of these materials can also be grown by vdW epitaxy [20, 21, 22].

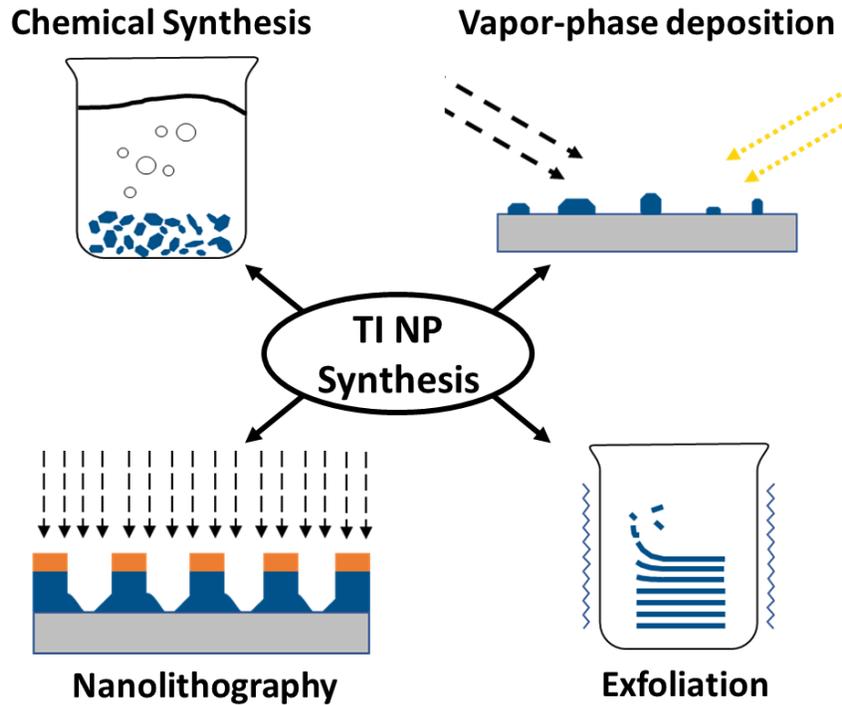

FIG. 2. Examples of nanomaterial synthesis techniques

Synthesis of TI nanoparticles, as well as nanoparticles of other 2D materials, has been reported via several methods as depicted in Figure 2 including nanolithography of thin films, vapor-phase growth via droplet epitaxy, solution-based synthesis, and exfoliation. We focus on 0D particles, in which the material is quantum-confined in all three dimensions. We do not consider nanowire synthesis, which has been discussed extensively elsewhere. In discussing each of these methods, we will focus on the degree of control over the nanoparticle dimensions, the nanoparticle polydispersity, the nanoparticle density, and the quality of the material. The dimensions of the nanoparticle control the energetic separation of the discrete electronic states, the polydispersity effects the variation in energy level spacing among the particles in an ensemble, and quality refers to the degree of crystallinity, the range of crystalline phases obtained, and the presence of defects/vacancies/other unwanted residues.

**A. Nanolithography of Thin Films**

The first method we discuss is a top-down technique: fabrication of thin films into nanostructures. The material is first confined in the thickness direction via the controlled growth of a thin film by molecular beam epitaxy (MBE), chemical vapor deposition (CVD) [17, 23], or other large-area deposition techniques. These techniques make use of van der Waals epitaxy, which eases the restrictions on viable substrate crystal structure and lattice constants. However, the presence of van der Waals bonding at the film-substrate interface does not inherently guarantee pristine, continuous thin films of the desired orientation. Several studies have been done on the optimization of growth parameters and how varying parameters and growth steps affect the final film properties. For instance, for 3D TI materials such as $Bi_2Se_3$, multiple works have been published on the growth of films on sapphire and GaAs, with factors such as the selenium overpressure and substrate temperature playing a major role on parameters such as the film orientation, domain size and the density of twinning defects [24, 25, 26].

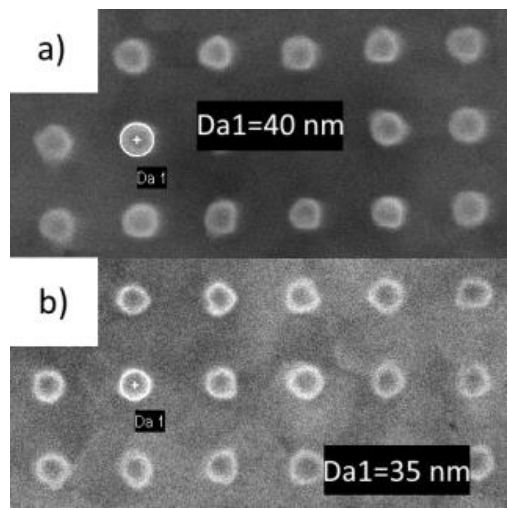

FIG. 3. SEM images of AR-N 7520.073 negative e-beam resist on $Bi_2Se_2$ post-development. a) and b) spun at different speeds prior to baking and written at different EBL doses. Resist in b) diluted with PGMEA solvent to reduce thickness. Reprinted from J. Vac. Sci. Technol. B 38, 055001 (2020) with permission from American Vacuum Society [27].

The other two degrees of confinement for the nanoparticle are provided via lithography. An etch mask can be patterned on the film via a photoresist combined with an electron beam [27], or by applying a monolayer of nanospheres [28] on the film surface, and the pattern is transferred into the film via an anisotropic etch. This approach offers an extremely high degree of control over the nanoparticle size and position, given that MBE- or CVD-grown films are highly uniform in thickness, and e-beam lithography offers control over the other two dimensions on the nanometer scale. We have fabricated nanoparticles via e-beam lithography in $Bi_2Se_3$ films for example with features 40 nm in diameter with 10 nm variation. An example SEM image of the nanoparticles (resist post-development) is shown in Figure 3. [27]. The ability to independently control particle dimensions and particle placement makes lithography extremely attractive, especially when compared to other approaches that produce samples with higher polydispersity and/or uncontrolled positions. However, lithography is an approach that inevitably results in exposure of the material to atmospheric contaminants, chemical developers, polymers, and solvents. The exposure to air is an immediate concern given that most vdW materials oxidize [29, 30, 31]. The presence of unwanted materials on any of the surfaces of the nanoparticles may act as an impediment to surface-sensitive probes and may introduce other conduction pathways, trap states, or additional sources of electrons that would make it difficult to observe the electronic states. Lastly, etch-induced damage on the outer surfaces of structures created by lithography may increase the rate of non-radiative recombination.

**B. Droplet epitaxy growth**

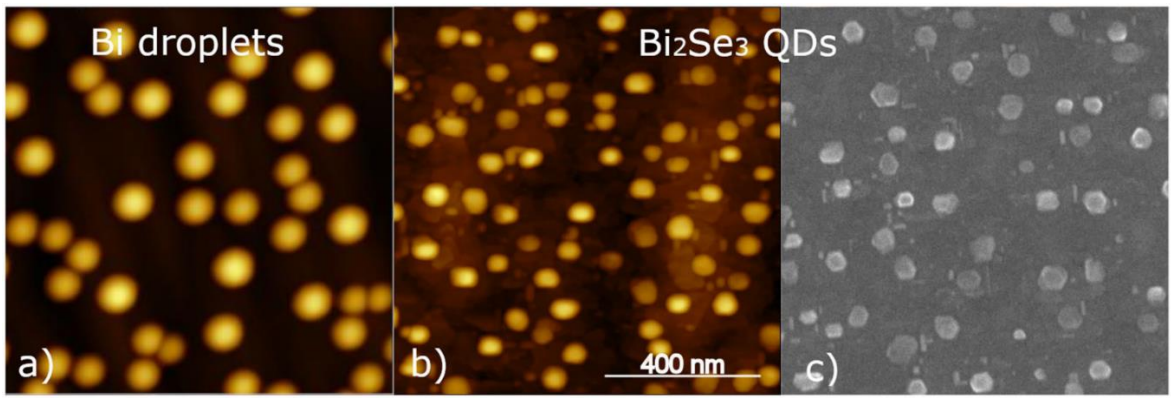

FIG. 4. Bi droplets (a) and Bi$_2$Se$_3$ nanoparticles in AFM (b) and SEM (c), all images share the same scale. Average nanoparticle height of 12 nm and width of 46 nm. Reprinted from Sci. Rep. 9, 3370 (2019), open access [32].

Another way to synthesize nanoparticles is through vapor-phase deposition. By controlling parameters of deposition, such as atomic flux and substrate temperature, the controlled nucleation of nanoparticles can take place. In III-V materials where the deposited material forms strong covalent bonds with the substrate surface, this can be achieved through strain-driven mechanisms such as Stranski-Krastanov growth [33, 34]. In the case of vdW materials however, this kind of covalent bonding and strain-driven growth does not occur due to the presence of the vdW gap. Therefore, an alternative technique known as droplet epitaxy is used for growing nanoparticles. In a vacuum deposition chamber, the metallic component, for instance Bi in Bi$_2$Se$_3$, is deposited using a low atomic flux for a short period of time. Due to the low wettability of the metal, it forms droplets. These droplets are then exposed to a flux of the second element in the material, for instance the chalcogen Se. These atoms impinge on the metal droplets, incorporate, and bond to form nanoparticles of the binary compound. Droplet epitaxy has been demonstrated for a range of systems, including the growth of semiconductor quantum dots such as InAs and GaAs [35, 36, 37]. In the case of TIs, this technique has been reported for MBE growth of Bi$_2$Se$_3$ nanoparticles on GaAs (001) (see example in Figure 4), with dimensions varying from 5-40 nm in height and 10-80 nm in diameter [32, 38]. Compared to nanolithography, droplet epitaxy is a much simpler process requiring only two deposition steps. Droplet epitaxy can give high densities of nanoparticles with dimensions as small as 10 nm, which is hard to achieve with e-beam lithography. The use of MBE also naturally results in the growth of extremely pure nanoparticles with a high degree of crystallinity. However, there are a few challenges with this technique. First, controlling the nanoparticle dimensions and the density independently is difficult. We often observe that as one growth parameter is varied, multiple nanoparticle properties change simultaneously. Next, the polydispersity of particles grown by this technique is high relative to nanofabricated samples. Finally, only GaAs has been explored as a substrate for the growth of TI nanoparticles. While it is likely that many vdW materials could be grown this way, further research is required to widen this technique's applicability.

**C. Solution-phase reaction synthesis**

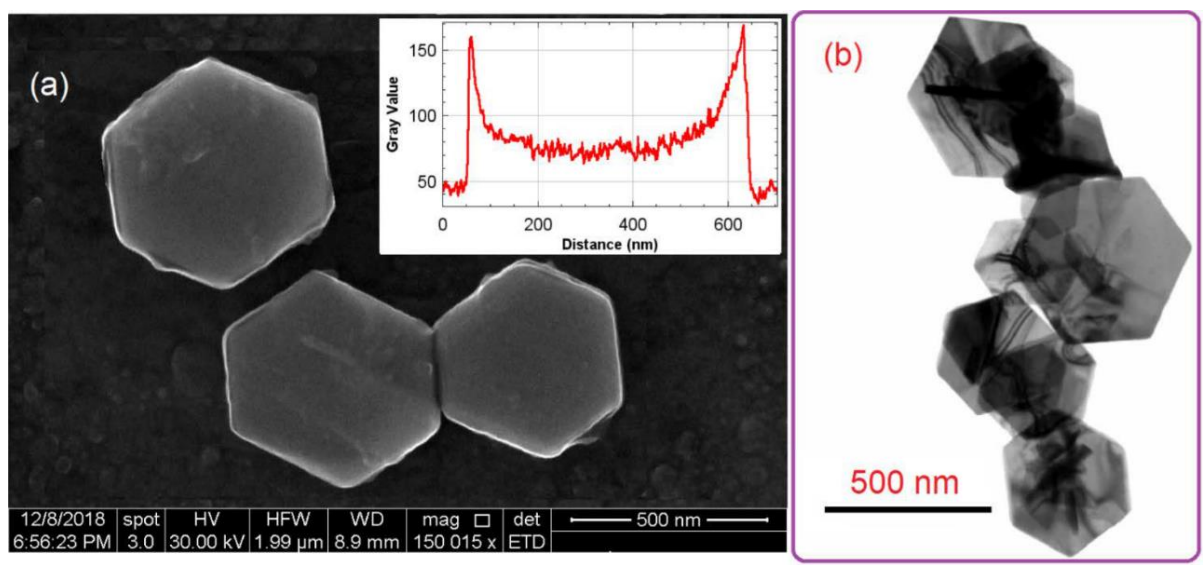

FIG. 5. Field-effect scanning electron microscopy image (a) and low-magnification TEM image of $Bi_2Se_3$ nanoparticles showing hexagonal morphology. Inset in (a) shows line profile in intensity for measurement of particle width. Reprinted from ACS Appl. Nano Mater. 3, pp 274-283 (2020), with permission from American Chemical Society [39].

Chemical synthesis of nanoparticles in aqueous solution is another method which has been demonstrated for growing nanostructures of various materials, including III-V semiconductors like InAs [40], 2D materials like $MoS_2$ [41], and TIs like $Bi_2Se_3$ and $Bi_2Te_3$ [42, 43, 39]. Sizes and geometries of these nanoparticles can vary greatly, from nanospheres as small as 10nm in diameter to nanoplates of widths as large as 600nm (Figure 5). The main advantage of this method is its ability to create a large volume of nanoparticles very cheaply. This technique can also create particles with spherical geometry, which is difficult to realize by other methods, and can be used to create core-shell geometries combining multiple materials [44, 45, 46]. Lastly, since the nanoparticles are freely dispersed in solution, they can be cast onto any substrate. However, synthesis of nanoparticles via solution reaction chemistry presents a few challenges. These nanoparticles do not always form in the desired crystal orientation and may contain impurities from precursor molecules or byproducts of the reaction. In addition, preparing nanoparticles for spectroscopic measurement from solution is not easy, as they typically need to be deposited onto a substrate in a specific orientation. This is difficult to guarantee, since most conventional methods of casting particles onto a substrate, such as drop-casting and spin coating, disperse particles in random orientations. The use of ligands to modify the particle surface and allow them to arrange into ordered arrays can potentially be employed to mitigate these disadvantages [47].

**D. Exfoliation**

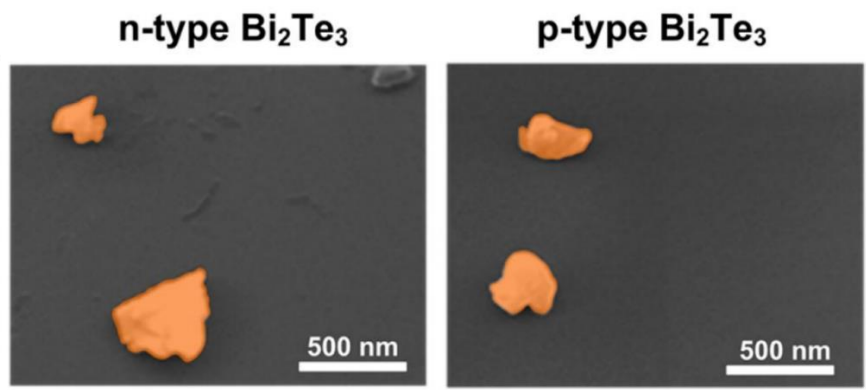

FIG. 6. SEM images of n-type and p-type $Bi_2Te_3$ nanoparticles created by grinding and trituration. Reprinted from ACS Photonics, 2, pp. 481-490 (2015), with permission from American Chemical Society [48].

The last approach we describe here is exfoliation. Exfoliation is a technique that exploits the unique vdW nature of 2D materials to create nanostructures. By employing mechanical or electrochemical stimuli, the vdW bonds between individual layers in a bulk crystal can be broken to produce nanoscale quantities of the material. In the case of TIs, groups have reported agitating bulk crystals in solvents [49], applying electrochemistry [50], and mechanical grinding and trituration [48]. The morphology of particles produced via these methods varies, with agitation and electrochemical synthesis tending to form nanosheets of few nanometers thickness and several hundred nanometers width, while grinding and trituration produce a crystallite morphology with a diameter of 200-300nm. We show an example of $Bi_2Te_3$ particles made via this latter method in Figure 6.

This technique shares many advantages with solution synthesis: exfoliation can produce a large volume of nanoparticles very cheaply, and the substrates these particles can be placed on are not as limited as in the case of vapor-phase growth. In addition, exfoliation starts with a bulk crystal, which may reduce the cost even further and limits the incorporation of precursor impurities. The accessibility of this type of synthesis is, however, its biggest advantage. Compared to ultra-high vacuum techniques, exfoliation can be done using simple methods such as grinding and sonication, which can be performed using widely available tools. While this simplicity and inexpensiveness are valuable for enabling scalable nanoparticle synthesis, the same detriments that were discussed for solution-based synthesis are further exacerbated in this case: high variance in nanoparticle dimensions and difficulty in controllably casting the nanoparticles onto the substrate to facilitate spectroscopy.

The variation in properties of the nanoparticle such as dimensions, density, polydispersity and quality as a function of the synthesis method is summarized in the following table:

TAB. 1. Summary of nanoparticle properties (dimensions, polydispersity, density and crystal quality) for each synthesis method

| | |
|---|---|
| Nanofabrication | Dimensions: As low as tens of nanometers (using EBL) |
| | Polydispersity: Tens of nanometers or lower (using EBL) |
| | Density: Highly controllable, but required EBL write time increases rapidly with increasing number of features |
| | Crystal quality: Dependent on the method used to create the initial bulk material being nanofabricated. Post-processing residues and etch-induced damage may have increased impact on material properties due to higher surface area-to-volume ratio in nanostructures |
| Droplet epitaxy | Dimensions: Between several nanometers and tens of nanometers

Polydispersity: Between several nanometers and tens of nanometers

Density: Can be affected by deposition time of initially nucleating metal (e,g, bismuth)

Crystal quality: High degree of crystallinity achievable with MBE |
| Chemical synthesis | Dimensions: Tends towards tens to hundreds of nanometers, with few monolayer thickness achievable

Polydispersity: Moderate to high polydispersity in nanoparticle dispersion, can be reduced by post-synthesis sieving to filter out particles by size

Density: Very high density achievable in relatively small volume of solution

Crystal quality: Highly crystalline nanoparticles, with wide range of geometry |
| Exfoliation | Dimensions: Generally closer to hundreds of nanometers, possible to achieve few layer nanosheets

Polydispersity: Relatively high polydispersity in particle size

Density: Extremely high, dependent on amount of initial source material

Crystal quality: Dependent on quality of initial bulk crystal. Generally very irregular geometry |

**III. CHARACTERIZATION OF NANOPARTICLE BAND STRUCTURE**

We will now discuss some of the methods available for probing the band structure of these nanoparticles. As nanoparticle dimensions are reduced to less than a few hundred nanometers, it becomes more difficult to create electrical contacts on these structures for transport studies. Therefore, we must employ alternative methods of probing these materials. We separate the discussion of these techniques into two groups: ensemble measurements and individual measurements. With ensemble techniques, the probe gathers information on many particles simultaneously, with the final measured quantity being a combination of the individual contributions of all the measured particles. Individual measurement involves using probes that are focused in an area that is of a similar size scale as a single particle and can therefore measure each nanoparticle independently. In the following sections, we will discuss angle-resolved photoemission spectroscopy (ARPES) and THz transmission spectroscopy for ensemble measurement, and scanning tunneling microscopy/spectroscopy (STM/STS) and electrostatic force microscopy (e-EFM) for individual measurement.

**A. Ensemble measurement**

In preparing a sample of many nanoparticles to be measured simultaneously, the goal is to maximize the density of particles and minimize the polydispersity in particle dimensions. Maximizing the surface coverage is essential for obtaining a sufficiently large signal when measuring over large areas. Variation in dimensions can result in inhomogeneity broadening in spectroscopic measurement, which "smears out" the discrete energy spacing between particles to the point where discrete levels are unable to be observed within the resolution of the detector. For this reason, techniques such as ARPES and THz transmission spectroscopy are most compatible with samples made via techniques such as lithography where size variation is simpler to control.

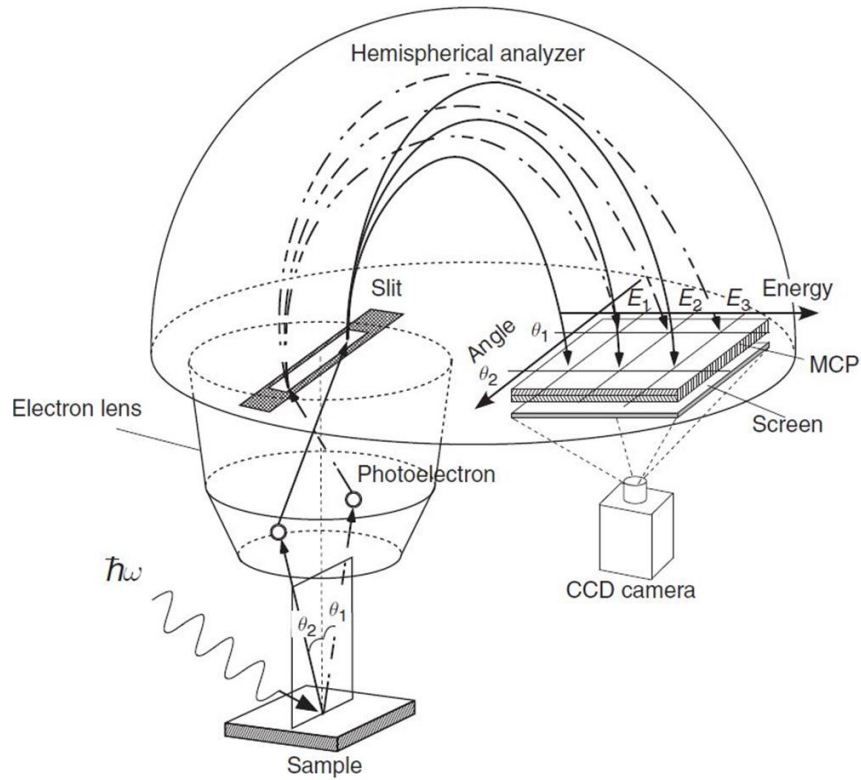

FIG. 7. Schematic of ARPES, Ejection of photoelectrons from sample surface, capturing of electrons through use of an electromagnetic lens and slit, propelling electrons through hemispherical analyzer and onto a multichannel plate (MCP). Reprinted

The first technique we discuss here is ARPES, and we show a schematic in Figure 7. ARPES typically involves shining a monochromatic light source on the surface of a material at an angle. This results in the ejection of electrons from the surface. These electrons are captured, typically by means of a hemispherical analyzer, and their kinetic energy is measured. The initial surface state binding energy of these electrons can be calculated using the initial energy of the light and the work function of the material. The electrons can also be captured as a function of the angle from which they were ejected, which can be used to calculate the in-plane and out-of-plane components of the ejected electron's wave vector. By making use of a system like the one in Figure 7, which propels the ejected electrons through a hemisphere and onto a multichannel plate (MCP), both the kinetic energy and polar angle of the ejected electron can be measured simultaneously. In ARPES, it is assumed that symmetry in the plane of the sample surface is unbroken in real space, and thus also in reciprocal space. As a result, the final in-plane wavevector of the ejected surface electron is equivalent to the initial in-plane wavevector of the energy state the electron was ejected from. From this, the surface band structure can be plotted.

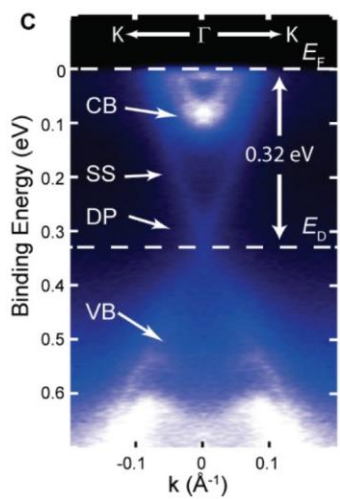

FIG. 8. ARPES data measured along the K-Γ-K direction for Bi$_2$Se$_3$ nanoplate aggregates grown on mica by vapor deposition. Bulk conduction and valence band edges as well as linearly dispersed surface states and Dirac point are indicated. Reprinted from J. Am. Chem. Soc. 134 (14) pp. 6132-6135 (2012), with permission from the American Chemical Society [52].

ARPES has been used to measure the band structure of buried semiconductor quantum dot systems like (In, Mn)As [53], as well as 2D materials such as graphene [54] and Bi$_2$Se$_3$ [52]. The primary advantage of this technique is the ability to resolve both the energy and wavevector of each state, something which is unique to this technique, and which allows for observation of the band dispersion for both surface and bulk bands. An example of ARPES conducted on Bi$_2$Se$_3$ nanoplates is shown in Figure 8 above. There are also modified versions of ARPES such as circular dichroism ARPES (CD-ARPES) that can resolve the spin of the electronic states, and this has been demonstrated for measuring the surface state spin texture in bulk Bi$_2$Se$_3$ crystals [55]. As both the dispersion and spin texture of the surface and bulk bands in TIs are different, techniques such as ARPES and CD-ARPES are well-suited for providing information on these states.

Because ARPES is a surface-sensitive technique, the major challenge is removal of atmospheric contaminants and oxidation through either pre-measurement baking or etching or cleaving the sample in vacuum to create a pristine surface. This latter method is not applicable for nanoparticles. While dry etching may have detrimental effects on the surface states [56], baking thin films and bulk crystals prior to measurement is effective. In thin films, proper tuning of the bake is necessary to avoid outgassing of elements like selenium which have a high vapor pressure, as this could result in degradation of the material. This issue is further exacerbated in the case of nanoparticles, which have a much higher surface area-to-volume ratio. Nanofabricated particles are also affected by the presence of additional processing contaminants. Overall, the surface sensitivity and simplicity of ARPES make it an effective technique at probing the surface band structure of thin films. Addressing the issue of pre-measurement surface treatment is needed to make this technique viable for measuring nanoparticles.

The next ensemble measurement method we discuss here is THz transmission spectroscopy. In this technique, light is shined on a sample and the transmitted signal is monitored as a function of wavelength. In the case of nanoparticles which exhibit quantum confinement, a drop in the transmitted signal may be correlated to absorption of the light at that wavelength due to excitation of electrons between discrete states. This is a relatively simple technique that has been used to measure the energy separation between discrete states in quantum dots like InGaAs [57] as well as intersubband transitions in GaN/AlGaN quantum wells [58].

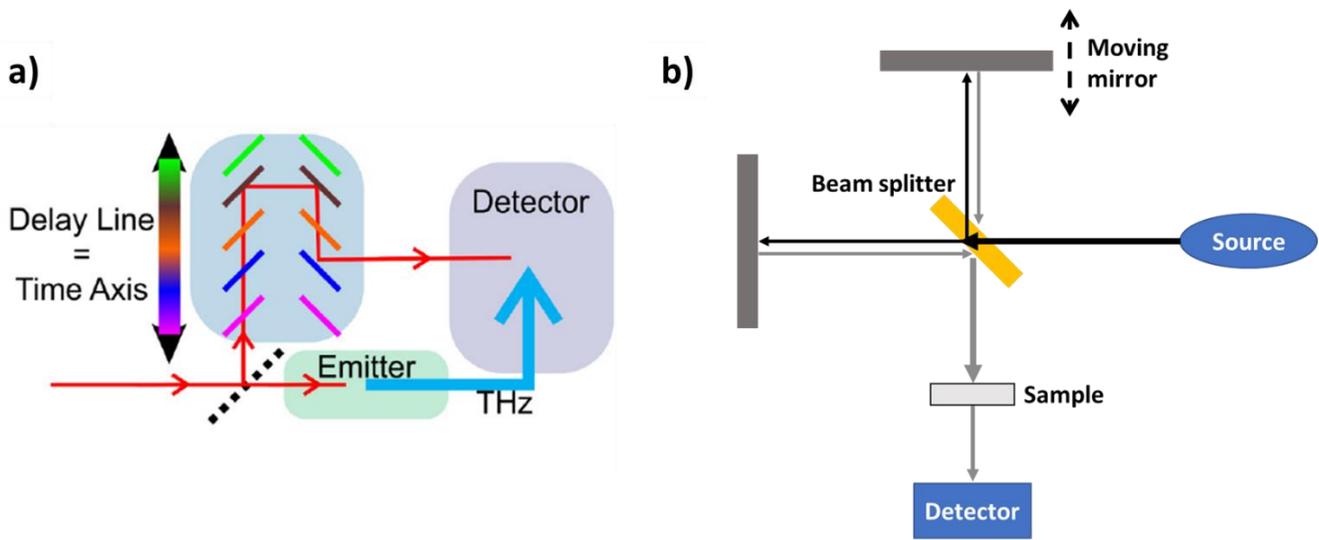

FIG. 9. Schematics of TDTS and FTIR. a) Simplified schematic of TDTS showing split beam and delay line to introduce a time delay. Reprinted from J. Appl. Phys. 124, 231101 (2018), with permission from AIP Publishing. b) Simplified schematic of FTIR showing Michelson interferometer, two beams created by a beamsplitter and one beam reflecting from a moving mirror to introduce a path difference.

THz light is well-suited for probing the energy spacings between discrete surface states in TI nanoparticles: based on reported theory, the discrete surface state spacing in nanoparticles of $Bi_2Se_3$ around ~100nm in size is anywhere from a few to several tens of meV, which in units of frequency is ~0-10 THz. Fourier Transform Infrared spectroscopy (FTIR) and time-domain THz spectroscopy (TDTS) are two widely used techniques for measuring THz transmission, and we show schematics of both techniques in Figure 9.

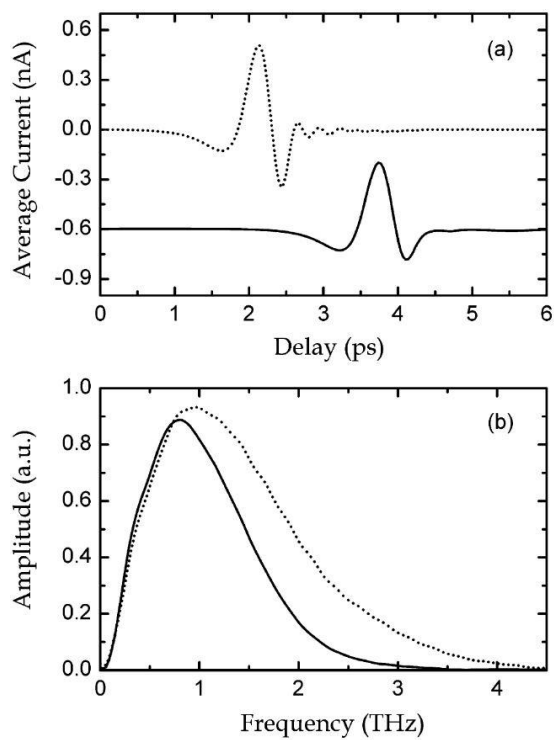

FIG. 10. THz Transmission data for ZnS nanoparticles synthesized by aqueous chemical reaction. a) shows measured THz pulses as a function of time and b) shows Fourier transform of time-domain data into the frequency domain. Dotted curves are for reference substrate, solid curves are for ZnS nanoparticles. Reprinted from J. Phys. Chem. B. 110(5), pp. 1989-1993 (2006), with permission from American Chemical Society [59]

Each technique has pros and cons: with FTIR, the THz source tends to be lower in intensity but covers a wider range of wavelengths, and simple FTIR systems can be purchased and used without modification. In TDTS [60], a THz laser pulse is used, which means the frequency range is much narrower, but the signal is much higher. TDTS setups can be more complicated and require frequent alignment to ensure the laser is focused on the sample. We show an example of transmission data in Figure 10 acquired for THz time-domain spectroscopy of ZnS nanoparticles mounted in a silicon cell over the range of 0.3 to 3 THz [59]. The general technique of measurement is to acquire data for the sample and a reference substrate (in this case an empty silicon cell) and correct the signal to the reference in order to determine where signal absorbed in the frequency domain due solely to the material of interest. In this case data is acquired by measuring the picosecond THz pulse over the time-domain and performing a Fourier transform to measure amplitude of the signal in the frequency domain.

While the simplicity of performing these measurements is a significant advantage, there are a few challenges in the interpretation of the measured data. First, because the goal is to correlate peaks in THz absorption to the transitions between energy states, we need to know the Fermi level in the material, as this controls the relative occupation of the initial and final states in the nanoparticle. We will elaborate on the issues related to controlling the Fermi level in the following section, but the standard approach of using a back gate is challenging for optical measurements since most back gates are metallic and therefore opaque. Another challenge is the presence of Fabry-Perot oscillations in the data. These oscillations occur when the light interferes with itself due to multiple orders of reflection within the sample. Deconvoluting these features from the real data can be difficult, as the absorption from the nanoparticle can be low compared to these Fabry-Perot oscillations. Overall, FTIR and TDTS are effective techniques that enable measurement of an ensemble of particles, but correlating the observed signal to the energy level spacing can be challenging. While deconvoluting artifacts from the data is important, the main challenge with this technique is achieving effective control over the Fermi level to ensure that excitation between discrete states can occur.

**B. Individual particle measurement**

We now move to discuss techniques which measure individual particles. We begin with STM/STS. This is a technique that makes use of an atomically sharp, conducting tip to measure the local electronic density of states. By fixing the location of the tip in the plane of the surface and applying a bias voltage, electrons can tunnel across the gap between the sample and tip. The variation in the tunneling current with bias voltage can be correlated to the local density of states at that point on the sample. The use of an atomically sharp tip allows this technique to have a high degree of spatial resolution in the x-y plane, making it promising for probing the discrete states in individual nanoparticles.

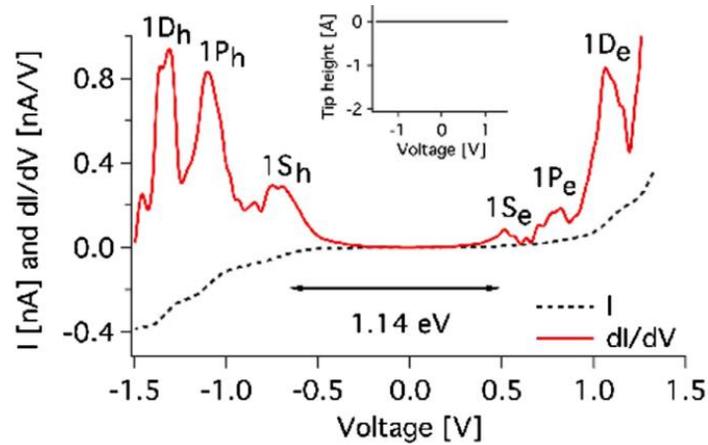

FIG. 11. Scanning tunneling spectroscopy data measured on PbS nanocrystals. Black dotted curve shows tunneling current, red solid curve shows differential tunneling current with respect to bias voltage. Reprinted figure from Phys. Rev. Lett. 100, 127406 (2013), with permission from American Physical Society [61].

We show an example of data acquired by this technique for PbS nanocrystals in Figure 11 [61]: in this figure the black dotted curve shows the tunneling current and the red solid curve shows the derivative of tunneling current with respect to bias voltage. This latter quantity can be correlated to the local density of states. In nanostructures where quantum confinement produces discretization in the bands, peaks in the differential tunneling current are seen at specific values of the bias as the local chemical potential shifts to enable tunneling into/out of different states.

The ability to resolve the density of states using this tip-based method enables individual particle measurement. This is useful in cases where the synthesis of the particles results in high polydispersity in dimensions: each particle can be examined independently, and having size variation gives one the ability to correlate changes in the measurement (i.e., the spacing between peaks in the differential tunneling current) to changes in the nanoparticle dimensions. However, much like with ARPES, this technique is extremely surface sensitive. Thus, it is just as necessary with STM/STS to ensure that the particle surface is kept pristine. Any synthesis method where there is exposure of the particles to atmosphere or chemicals has the possibility of leaving residues on the surface of the particles which may impede tip-based probes. For this reason, nanofabricated particles are a relatively poor candidate for measurement by this method. One option is to use MBE to grow nanoparticles by droplet epitaxy, and then transfer the sample into an STM without breaking vacuum. This is a relatively simple process if the STM is already connected to the MBE but can also be achieved through use of a glove box/glove bag during unloading and loading, or with a vacuum suitcase.

Another tip-based technique for probing surface band structure is e-EFM [62]. Like STM, a bias is applied between a tip and sample to induce movement of electrons into/out of the nanoparticles on the substrate surface. However, unlike STM, the tip is positioned far enough from the surface such that electrons do not tunnel between the tip and the particle. Instead, electrons travel between the particle and a back-gate within the substrate that is used for biasing. The change in the overall charge within the particle is sensed by the tip as a change in the electrostatic interaction, causing a dissipation in its natural oscillation frequency.

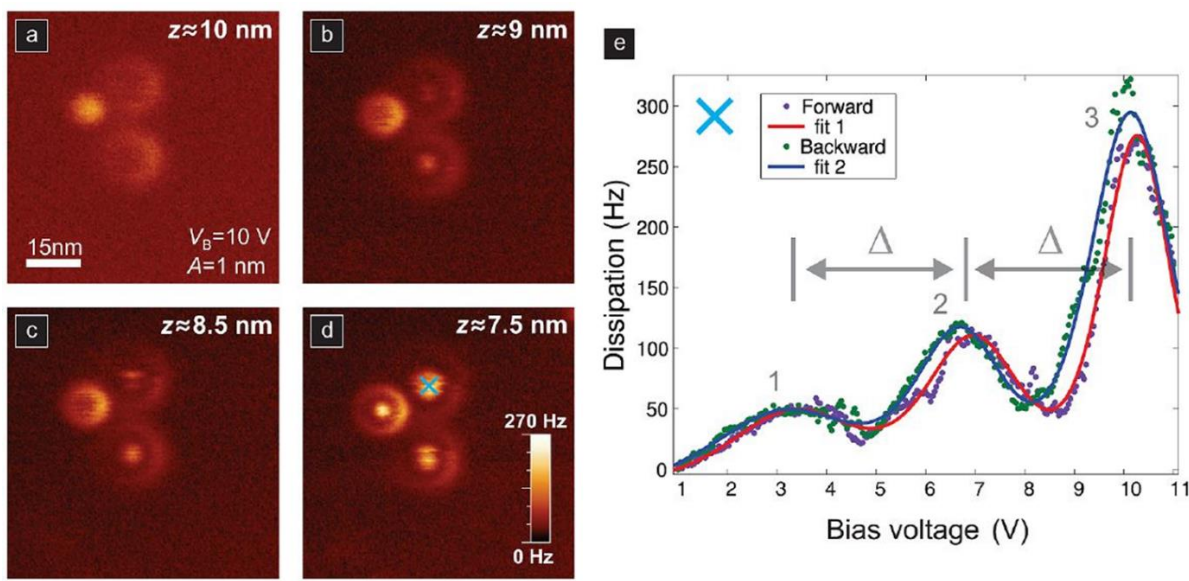

FIG. 12. Dissipation data measured by e-EFM on gold nanoparticles. a)-d) show sweeps of the tip measuring dissipation at a fixed tip-sample distance. e) shows peaks in dissipation as a function of bias voltage at a fixed tip position. Reprinted from ACS Nano 7, 5, 4683–4690 (2013), with permission from American Chemical Society [63].

This technique has been applied to probe several 2D materials including graphene [64, 65] and WSe$_2$ [66]. These studies show that e-EFM can detect changes in the electrical response associated with a change in thickness of singular monolayers to enable the measurement of nanoscale changes in height and dielectric constant. e-EFM has also been applied to several nanoparticle systems such as InP [67] and Au [63], specifically to understand the band structure. We show examples in Figure 12 of data acquired by e-EFM for Au nanoparticles on NaCl monolayers deposited on a Fe substrate. Figure 12 a)-d) show data acquired as the tip sweeps over the surface with a fixed bias controlled by fixing the height of the tip to *z*. As the tip moves, the bias experienced by the particle changes. At tip positions where the bias results in motion of charge from a specific discrete state in the particle, a ring appears in the dissipation around the particle. Figure 12 e) shows data acquired when the tip is kept at a constant location over a particle and the bias voltage is varied. Peaks appear in dissipation at specific bias voltages associated with charge tunneling into/out of a discrete state in the particle.

e-EFM has many of the same advantages of STM/STS, given its ability to measure individual particles on the surface. However, because this technique does not measure transport between the surface and the tip, the requirement for surface cleanliness is less stringent. For instance, the use of capping layers to protect particles prior to STM/STS measurement can reduce the possibility of oxidation. However, these layers may inhibit electron transport and thus may need to be removed prior to measurement. However, if there is no concern about the capping layer interfering with the electrostatic interaction between the tip and the particles underneath the cap, then this de-capping procedure may be unnecessary for e-EFM. One of the main challenges with e-EFM is interpreting the data. The peak-to-peak spacing as observed in dissipation vs. bias plots requires both modelling and an understanding of the band alignment between the particles and the substrate, so that the dissipation can be attributed to charge transferring from the specific discrete states of interest (for instance the discrete surface states in the TIs). For many vdW materials, there is very little data available on the band alignment between the material and typical semiconductors. Overcoming this challenge will involve either obtaining band alignment information or through careful design of the substrate and back-gate such that the characteristics of the charge-transfer process under a bias are well understood.

## IV. KEY CHALLENGES IN NANOPARTICLE MEASUREMENT

While we have routes to both synthesize and characterize nanoparticles, several key challenges to probing their quantum-confined states remain. We will highlight a few of these here and offer some ideas as to how these issues can be overcome.

### A. Controlling Fermi level

In general, methods for characterizing the electronic band structure of a material require a degree of understanding and some control over which states are occupied. In the case of ARPES, electrons are only photoemitted from occupied states: the Fermi level must therefore be above the states of interest. THz absorption spectroscopy requires the initial state to be occupied and the final state to be unoccupied in the absorbing transition. For example, if the Fermi level is either near or above (below) the bulk conduction (valence) band edge in a TI, then the surface states which exist within the bulk band gap will be completely occupied (unoccupied). This means that we would be unable to observe transitions of electrons between surface states under THz light. Adjusting the occupation of electronic states is often achieved by using a top or back gate. However, as was mentioned earlier, the use of a back-gate may not be compatible with THz transmission measurements. An alternative approach is ion gel top-gating, which has been used to gate graphene and carbon nanotubes for THz absorption studies [68, 69]. Further work is required to optimize this process for vdW nanoparticles to ensure equal gating across the ensemble and to maximize gate transparency in the frequency range of interest.

Controlling the Fermi level can also be achieved at the synthesis stage. Composition control is one option. This may involve adjusting the relative atomic ratio in an alloy, or introducing dopants that compensate for excess electrons or holes. For example, in TIs, adjusting the Bi to Sb ratio in the ternary compound $(Bi_xSb_{1-x})_2Te_3$ results in a shift from n-type conductivity to p-type conductivity [70]. Dopants, like Ca in $Bi_2Se_3$, have also been used to tune the Fermi energy during growth [71]. We note that this method of adjusting composition or doping is more compatible with certain synthesis methods than others. Doping and varying compositions of thin films and bulk crystals for the purpose of exfoliation or nanofabrication is more straightforward than trying to control the incorporation of dopants and multiple alloying elements in reaction chemistry or droplet epitaxy growth. Despite the usefulness of these methods to adjust the Fermi energy during growth, ultimately varying the Fermi level during the measurement itself is likely a better option since this allows multiple measurements on a single particle to be conducted with a varying bias.

**B. Chemical sensitivity**

A substantial problem in characterization of 2D materials is air sensitivity, as these materials tend to oxidize rapidly [30, 29, 72]. In nanostructures, this poses a significant problem due to the increased surface area to volume ratio. Surface oxidation can cause a reduction in signal by creating an additional barrier between the probe and the nanoparticle. One way to preserve the pristine surface is through capping layers. For many vdW materials, amorphous or crystalline chalcogenide capping layers have been grown on thin films to protect the surface from oxidation and organic contaminants after removal into air. Prior to measurement in XPS, ARPES and/or STM, the chalcogenide cap is thermally desorbed [56, 73]. However, the tuning of this pre-measurement bake is a sensitive process due to the relatively high vapor pressure of the chalcogenides; overbaking may result in depletion of the chalcogen from the film [31]. This poses a substantial risk to nanostructures. Further work to investigate ideal capping and baking procedures for nanoparticles, or identifying alternative capping layers and de-capping methods besides heating would be useful. As mentioned earlier, making use of vacuum suitcases or glove boxes for packaging samples so that exposure to air is minimized is another viable option.

**C. Determining band alignment**

The last point on which further work is needed is determining the band alignment between the nanostructures and substrates or gating materials. In discussing techniques like STM/STS and e-EFM, the substrates used for performing these measurements may include many common semiconductors due to their ease of preparation for growth, relatively low cost, and their ability to act as gates. However, information regarding the band alignment between many common semiconductors and vdW materials is limited. Definitive attribution of discrete peaks in tunneling current or dissipation to specific discrete states in the nanoparticle requires rigorous modelling and a clear understanding of the band alignment. Gaining this information is crucial to designing samples and interpreting data acquired by STM/STS and e-EFM.

**V. CONCLUSION**

The breadth of research into the physics and applications of nanostructures of vdW materials has grown rapidly over the past decade. Multiple innovations in the synthesis and characterization of nanomaterials have shown that the unique physics of materials under quantum confinement has the potential to revolutionize many industries. However, the work that has been done has also made clear the challenges that come when working with material at the nanometer size scale. The next goal of researchers in this area will be to overcome the challenges of sample preservation and characterization in order to better understand how to make these materials more robust and stable, and thus improve the feasibility of incorporating quantum-confined materials into more commercial and consumer devices. Making quantum-confined materials more widely useable massively improves the efficiency of space for components on chip and by extension can improve processing speed and power and could enable quantum computers to transition from being large industrial installations into seeing use in readily available computing devices.

## ACKNOWLEDGMENTS

S. V. M. and S. L. acknowledge funding from the National Science Foundation, Division of Materials Research under Award No. 1838504 and the Brookhaven National Laboratory/University of Delaware Seed Program under Award No. 20A00145. This material is based upon work supported by the University of Delaware Graduate College through the University Doctoral Fellowship Award. Any opinions, findings, and conclusions or recommendations expressed in this material are those of the author(s).

## AUTHOR DECLARATIONS

The authors have no conflicts to disclose.

## DATA AVAILABILITY STATEMENT

Data sharing is not applicable to this article as no new data were created or analyzed in this study.